\documentclass[twocolumn]{aastex701}
\usepackage{graphicx}
\usepackage{amsmath}
\usepackage{amssymb}
\usepackage{subcaption}
\usepackage[export]{adjustbox}

\begin{document}

\title{Delayed Radio Flares in Neutrino-associated Blazars: The Case of TXS 0506+056}
\correspondingauthor{S. I. Stathopoulos}

\author[orcid=0000-0002-1344-3754]{S. I. Stathopoulos}
\affiliation{Deutsches Elektronen-Synchrotron DESY, Platanenallee 6, 15738 Zeuthen, Germany}
\email[show]{stamatios.ilias.stathopoulos@desy.de}

\author[orcid=0000-0003-0327-6136]{C. Yuan}
\affiliation{Université Libre de Bruxelles, CP225 Boulevard du Triomphe, 1050 Brussels, Belgium}
\email{chengchao.yuan@ulb.be}

\author[orcid=0000-0003-3902-3915]{G. Vasilopoulos}
\affiliation{Department of Physics, National and Kapodistrian University of Athens, University Campus Zografos, GR 15784, Athens, Greece}
\email{gevas@phys.uoa.gr}

\author[orcid=0009-0000-9401-1971]{F. Testagrossa}
\affiliation{Deutsches Elektronen-Synchrotron DESY, Platanenallee 6, 15738 Zeuthen, Germany}
\email{federico.testagrossa@desy.de}

\author[orcid=0000-0002-2383-3860]{D. Karavola}
\affiliation{Department of Physics, National and Kapodistrian University of Athens, University Campus Zografos, GR 15784, Athens, Greece}
\email{dkaravola@phys.uoa.gr}

\author[orcid=0000-0001-6640-0179]{M. Petropoulou}
\affiliation{Department of Physics, National and Kapodistrian University of Athens, University Campus Zografos, GR 15784, Athens, Greece}
\email{mpetropo@phys.uoa.gr}

\author[orcid=0000-0001-7062-0289]{W. Winter}
\affiliation{Deutsches Elektronen-Synchrotron DESY, Platanenallee 6, 15738 Zeuthen, Germany}
\email{walter.winter@desy.de}

\begin{abstract}
Radio flares have been postulated to be associated with the production of astrophysical neutrinos. For example, TXS 0506+056 exhibits a 2–3 yr delay between the 2017 IceCube-170922A/$\gamma$-ray flare and a GHz radio maximum. We quantitatively test if the delayed radio flare originates from the same compact region where neutrinos and $\gamma$-rays are produced as it expands downstream and synchrotron self-absorption (SSA) is reduced. Starting from the 2017 flare blob parameters, we model the expanding production region and its evolving radio emission with {\tt LeHaMoC} in a fully time-dependent framework, and compare our 1.2–22 GHz light curves to RATAN-600 data. We study different scenarios with increasing levels of sophistication, including continuous injection and energy re-dissipation on pc scales. While a simple expanding blob scenario fails to reproduce the radio data, a downstream dissipation episode of particles in the optically thin regime, followed by jet deceleration, successfully describes the radio evolution. The radio flare is powered by leptonic synchrotron emission and is largely insensitive to the proton population relevant for neutrino production, implying that the delayed radio flare mainly probes downstream dissipation and beaming in certain jet configurations instead of being a direct radiative counterpart of the compact neutrino-production zone.
\end{abstract}


\section{Introduction}
High-energy astrophysical neutrinos ($E_{\rm \nu}>10$ TeV) probe hadronic processes in relativistic jets, but the location and properties of their production sites remain uncertain.  A milestone real-time association between a high-energy neutrino and an active galaxy came with IceCube-170922A, detected in September 2017 during a bright $\gamma$-ray flare from the blazar TXS~0506+056\footnote{Earlier individual AGN-neutrino associations had been proposed, for example the temporal and positional coincidence between the PeV event IC35 and the blazar PKS~B1424--418 \citep{2016NatPh..12..807K}.} \citep{2018Sci...361.1378I}. Multi-wavelength observations from radio to very high energy $\gamma$-rays established TXS 0506+056 as a bright, variable masquerading BL Lac \citep{2019MNRAS.484L.104P} and enabled detailed leptohadronic modeling of the flaring state \citep{Keivani2018ApJ,2018ApJ...863L..10A,2019NatAs...3...88G,2019MNRAS.483L..12C, SOPRANO, 2022ApJ...927..197A}. Very-long-baseline interferometry (VLBI) observations revealed rapid expansion of the compact radio core after the neutrino detection \citep{2020A&A...633L...1R}. These findings identified blazar jets as promising neutrino factories, but the connection between inner-jet activity and longer-term parsec-scale radio emission remains unclear.

Subsequent detections have strengthened the case for blazars as multi-messenger sources \citep[e.g.,][]{2016MNRAS.457.3582P, 2018Sci...361.1378I, Giommi2020, 2020PhRvL.124e1103A}, while also revealing diverse temporal behavior. In TXS~0506+056, the IceCube-170922A neutrino arrived during a major $\gamma$-ray flare, while RATAN-600 monitoring shows that the associated GHz radio outburst started rising around the same epoch but peaked only around $\sim$2020, implying a $\gamma$–radio delay of 2–3 years, and with lower radio frequencies peaking even later \citep{2024MNRAS.527.8784A}. The multi-frequency radio light curves exhibit a year-scale outburst envelope, with some frequency dependency. A second $>200$ TeV neutrino detected by Baikal-GVD in 2021 again coincided with the onset of a strong, multi-year radio flare in the same source \citep{2024MNRAS.527.8784A}. By contrast, in the quasar PKS~0446+11, the IceCube-240105A event occurred during a broadband $\gamma$-to-radio flare where all bands brightened nearly simultaneously and the radio peak lagged the $\gamma$-ray activity by at most a few months, well within $\sim 1$~yr from the neutrino arrival \citep{2025arXiv251107535K}. Polarimetric VLBI studies of PKS~0735+178 provide another example in which a neutrino-coincident multi-band flare was associated with structural and polarization changes in the parsec-scale jet, interpreted as a propagating shock front from the core region \citep{2025ApJ...989..208P}. Together, these sources illustrate that neutrino-associated blazar activity can range from long, multi-year $\gamma$–radio delays to nearly contemporaneous multi-band flares. Note that the idea that neutrino associations are related to enhanced radio activity has also been proposed in empirical or observational studies \citep{2020ApJ...894..101P,2021A&A...650A..83H,2024ApJ...964....3A}.

If the neutrino and $\gamma$-ray emission originates in the compact VLBI-core region, the delayed radio emission may arise if the same disturbance is initially opaque to GHz synchrotron radiation. As this region is advected downstream along the jet, it expands, its magnetic field decreases, and the SSA turnover frequency $\nu_{\rm ssa}$ shifts to lower values \citep{1966Natur.211.1131V,1985ApJ...298..114M}. In such "expanding blob" or shock-in-jet scenarios, a delayed radio flare is expected when $\nu_{\rm ssa}$ passes through the GHz band in a leptonic model \citep{2022A&A...657A..20B,2022A&A...658A.173T}, or when a proton population is present in the emitting region \citep{2023A&A...669A.151Z}, with higher frequencies peaking earlier than lower ones and the optically thick spectrum approaching $F_{\nu}\propto \nu ^{5/2}$ \citep{1986rpa..book.....R}. Time-dependent lepto-hadronic models for TXS~0506+056 already reproduce the multi-wavelength spectral energy distribution (SED) and candidate neutrino emission during the 2017 flare, constraining the size, magnetic field, Doppler factor, and particle content of a compact emitting region in the jet \citep{2018ApJ...863L..10A,2019NatAs...3...88G,2019MNRAS.483L..12C,2022ApJ...927..197A}. Combined with long-term radio monitoring, theoretical models allow a quantitative test of whether the delayed radio outburst can be explained as a simple expanding, self-absorbed evolution of the neutrino-emitting region.

In this Letter, we investigate for the first time the physical connection between the neutrino alert and the radio-flare observed in TXS~0506+056. Assuming that the neutrino and $\gamma $-ray emission during the 2017 flare originated in the same compact region, we test whether its evolution can describe the 2020 radio outburst. To do so, we consider three scenarios for the subsequent evolution of the emitting region and compare their predicted radio lightcurves with RATAN-600 data. We show that SSA-driven expanding blob models cannot reproduce the timing and evolution of the radio-flare, whereas a scenario invoking energy dissipation on parsec scales and jet deceleration provides a good description. TXS~0506+056 thus provides a case study in which the disturbance associated with the 2017 $\gamma$-ray flare, coincident with IceCube-170922A neutrino, later produces a delayed, optically thin radio signature downstream. This points to a dynamical connection to the 2017 event, but not a one-to-one radiative connection between the neutrino-emitting region and the delayed GHz outburst.

\section{Radio Data} 

\begin{figure}[ht!]
    \centering {\includegraphics [width=0.45\textwidth]{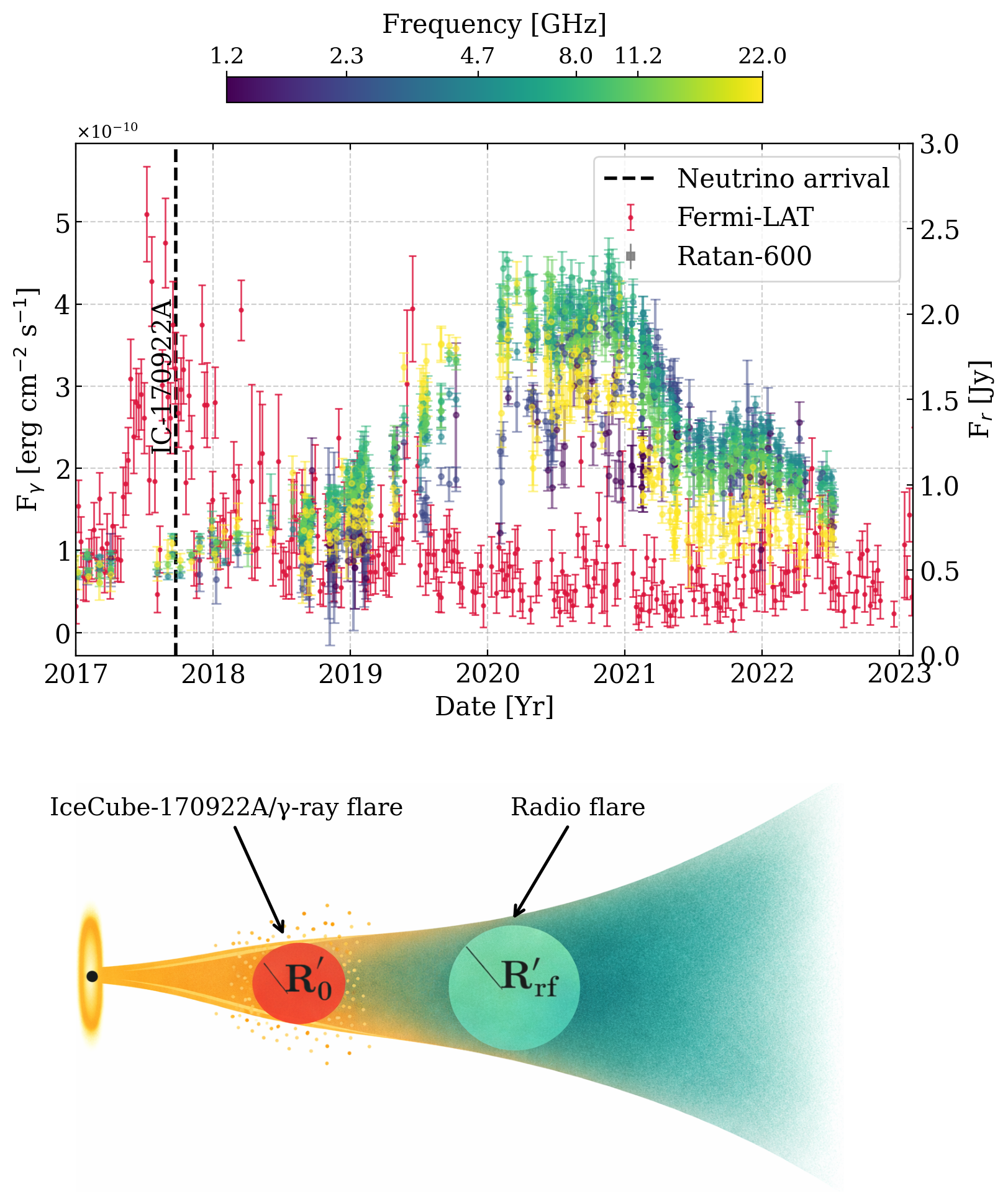}}
     \caption{Multiwavelength behavior and schematic model. Top panel: Long-term light curve showing the weekly Fermi-LAT $\gamma$-ray flux from the Fermi-LAT Light Curve Repository \citep{2023ApJS..265...31A}, based on the LAT likelihood analysis described therein together with the radio flux densities (points colored by observing frequency). The vertical dashed line marks the time of the high-energy neutrino arrival. Bottom panel: Schematic of the expanding blob model. A compact emitting region of initial radius $R'_0$ corresponding to the neutrino/$\gamma$- ray flare conditions inferred by \citep{Keivani2018ApJ}, propagates along the jet and expands. The outer radius $R'_{\rm rf}$ corresponds to the optically thin radio-emitting stage constrained by RATAN-600 and MOJAVE observations.}
      \label{fig:sigma_vs_p}
\end{figure}

For the long-term radio behavior of TXS~0506+056, we make use of the RATAN-600 light curves at 1.2–22 GHz covering 2010.0–2022.5, as presented in \cite{2024MNRAS.527.8784A}. RATAN-600 monitored TXS~0506+056 at six frequencies (1.2, 2.3, 4.7, 8, 11.2, and 22 GHz) and the resulting light curves, shown in their Fig. 5, were obtained with a consistent flux-density scale based on standard calibrators. In the analysis below, we focus on the interval around the 2020 outburst, during which all six RATAN-600 bands show a pronounced, common year-scale radio outburst envelope. (see top panel Fig.~\ref{fig:sigma_vs_p}). To also constrain the structure and size of the emitting region on parsec scales, we further use 15 GHz VLBA images of TXS~0506+056 from the MOJAVE program. Publicly available total-intensity images and movies for this source (B1950 name 0506+056) are accessible on the MOJAVE website\footnote{\url{https://www.cv.nrao.edu/MOJAVE/}}.

\section{Model} 

\begin{figure}
\centering
\includegraphics[width=\columnwidth]{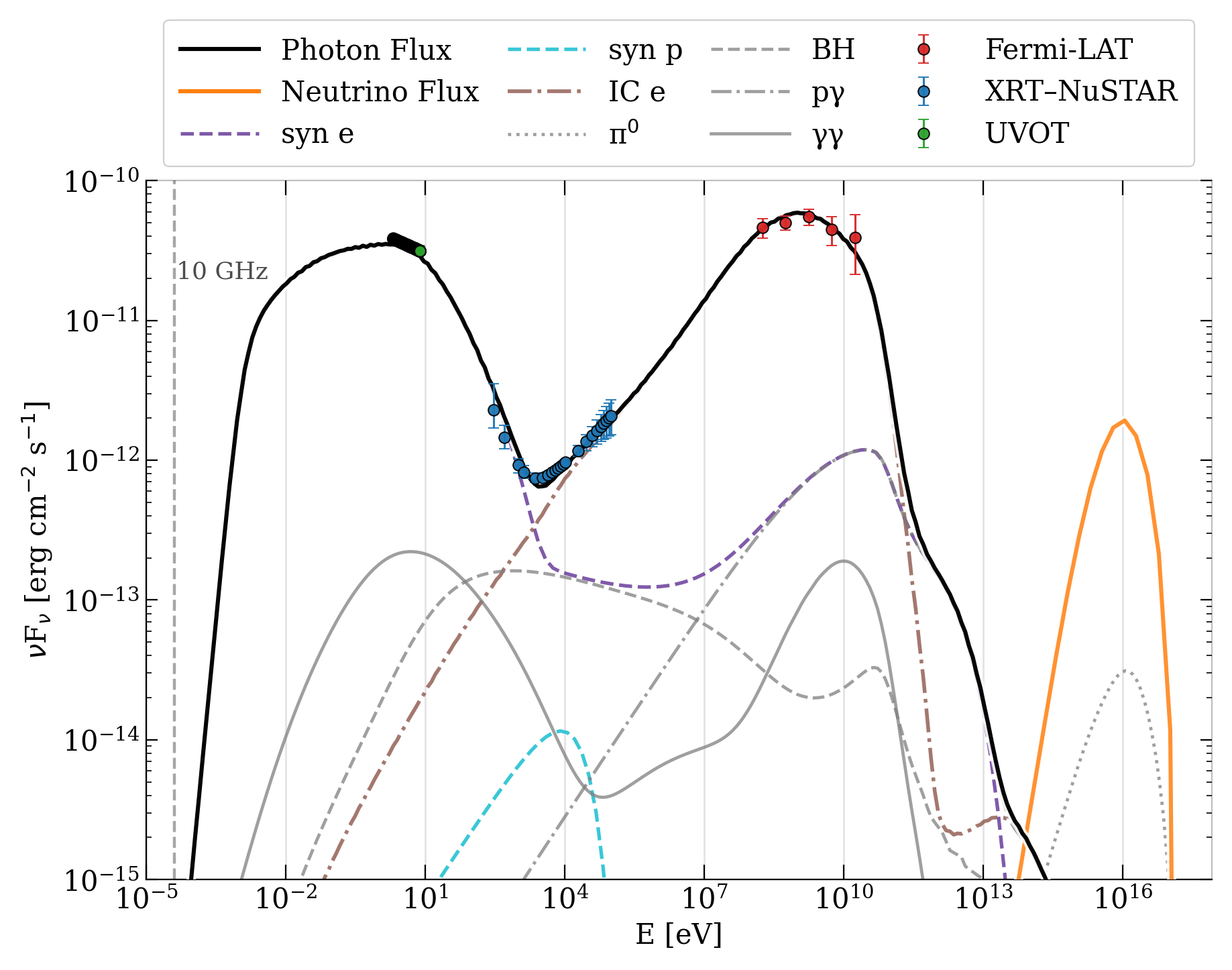}
\caption{SED at the neutrino epoch computed with {\tt LeHaMoC}, using the \citep{Keivani2018ApJ} LMBB2b leptohadronic solution. Different line styles represent different photon-production processes, grey curves denote secondary/hadronic components, while colored curves highlight the main leptonic contributions. The thick black solid line shows the overall photon spectrum, while the thick orange solid line shows the all-flavor neutrino spectrum.}
\label{fig:TXS_models_g_nu}
\end{figure}

The starting point of our analysis is the 2017 flaring state of TXS~0506+056, temporally associated with the IceCube-170922A neutrino and a bright GeV $\gamma$-ray flare. Modeling of the SED during that epoch can constrain the size of the emitting zone, its magnetic field, Doppler factor, and the distributions of relativistic electrons and protons. For the physical conditions during this episode, we adopt the parameters from the LMBB2b model of \cite{Keivani2018ApJ}, which yields the largest integrated all-flavor neutrino flux in the energy band of 0.1-10 PeV among their viable solutions\footnote{These parameters are broadly consistent with typical values inferred from one-zone modeling of the 2017 flare (e.g. $\delta_0 \sim 20$-$30$ and $R'\sim10^{16}$-$10^{17}\,\mathrm{cm}$ \citep{2019NatAs...3...88G,2019MNRAS.483L..12C}.}.
For completeness, in Fig.~\ref{fig:TXS_models_g_nu}, we show the full SED of the LMBB2b model, recalculated with {\tt LeHaMoC} \citep{2024A&A...683A.225S}, a time-dependent leptohadronic code that evolves the coupled kinetic equations of photons and relativistic particles (electrons, positrons, protons, and neutrinos) in a homogeneous magnetized region that may also be expanding. In all viable solutions of \cite{Keivani2018ApJ}, the synchrotron and inverse-Compton (IC) emission from primary electrons/positrons dominates the IR–GeV output, while the predicted GHz flux is strongly suppressed by SSA, indicating that a less compact region is responsible for the radio emission. We also note that the contribution of secondary pairs from hadronic cascades remains subdominant (see also \cite{2019NatAs...3...88G}). Moreover, any additional injection of secondary $e^\pm$ pairs via photohadronic (or Bethe-Heitler) interactions is expected to be most efficient in the compact flare zone where the target photon density is highest, and to decrease rapidly as the disturbance propagates and expands downstream. We therefore neglect the contribution of secondary pairs in the post-flare evolution and follow only the primary electron/positron population, using $B'_0$, $R'_0$, and $\delta_0$ (Table~\ref{tab:blob_params}) as initial conditions, where the primed symbols denote the quantities measured in the blob comoving frame. In the following, the same code is used to compute the time-dependent radio light curves for comparison with the observations.

\begin{deluxetable*}{lll}
\tablecaption{Model parameters used for the TXS~0506+056 radio-flare modeling.\label{tab:blob_params}}
\tablehead{
\colhead{Parameter} & \colhead{Value} & \colhead{Description}
}
\startdata
$B'_0$ [G]                    & 0.4                & Magnetic-field strength \\
$R'_0$ [cm]                   & $10^{17}$          & Radius of the emitting region \\
$\delta_0$                    & 24.2               & Doppler factor of the emitting region \\
$L'_{\rm e,0}$ [erg s$^{-1}$] & $2.2\times10^{42}$ & Injected electron/positron luminosity \\
$s_{e,1}$                     & 1.9                & Electron/positron spectral index below the break \\
$s_{e,2}$                     & 3.6                & Electron/positron spectral index above the break \\
$\gamma'_{\rm e,min}$         & 1                  & Minimum e$^{\pm}$ Lorentz factor \\
$\gamma'_{\rm e,b}$           & $5\times10^{3}$    & Break e$^{\pm}$ Lorentz factor \\
$\gamma'_{\rm e,max}$         & $8\times10^{4}$    & Maximum e$^{\pm}$ Lorentz factor \\
\tableline
$V_{\rm exp}~[c]$                 & $0.28$            & Radial expansion velocity in the blob comoving frame \\
$m_{\rm b}$                   & 0.56               & Magnetic-field decay index \\
$m_{\rm l}$                   & $-0.68$            & Electron/positron injection evolution index \\
$t'_{\rm reacc}~[R'_0/c]$              & $172.93$   & Onset of the re-dissipation episode \\
\enddata
\tablenotetext{}{Parameters above the horizontal line define the initial 2017 flare state adopted from \citet{Keivani2018ApJ}; those below describe the subsequent evolution in Scenario~C.}
\end{deluxetable*}

To initialize the expanding region, we inject a broken power-law distribution for the combined radiating electron/positron population with the same spectral shape as in the flare model (Table~\ref{tab:blob_params}) and evolve it at fixed $(R'_0,B'_0,\delta_0)$ until a quasi-steady (cooled) distribution is reached. The resulting electron/positron distribution that occupies the spherical blob $N'_{\rm e}(\gamma')$ is then adopted directly as the initial condition for the subsequent expansion phase. We model the emitting region as a spherical blob that expands with comoving velocity $V_{\rm exp}$, while the magnetic field decays as a power-law with radius, $B'(R')\propto R'^{-m_{\rm b}}$. When additional particle injection is included, the non-thermal electron/positron injection power is parameterised as $L_{\rm e}'(R')\propto R'^{-m_{\rm l}}$ to mimic a gradual evolution of dissipation efficiency as the disturbance propagates and expands. In Scenario A \& B, we consider expansion of the initial electron/positron population with and without further injection of particles (Sect. \ref{sub_sec_AB}), which aims to test whether the transition from optically thick to thin during the expansion of the blob can explain the delayed radio flare. We treat $V_{\rm exp}, m_{\rm b}$ and $m_{\rm l}$ (when applicable) as free parameters constrained only by the radio light curves and spectra. In a third scenario (Sect.~\ref{sub_sec_C}), where we explore a re-dissipation of energy episode in the optically thin regime, we additionally constrain $V_{\rm exp}$ with an independent estimate from the apparent expansion of the VLBI core in 15 GHz MOJAVE images of TXS~0506+056. The detailed derivation of this expansion speed and the associated uncertainties is described in Appendix~\ref{Appb}.

\section{Results}\label{Sec_Res}

Before presenting model comparisons, we note two key features of the RATAN-600 light curves around the 2020 outburst: First, the peaks at 1.2-22 GHz form a common year-scale radio-flare envelope. This puts an SSA-driven explanation to the test, since an SSA-dominated expanding-source scenario would generally predict that higher-frequency bands peak earlier as the SSA turnover frequency moves downward through the radio regime. Second, the post-peak decay is rapid and broadly similar across frequencies. Any viable scenario must therefore explain both the weak frequency ordering of the peak times and the fast decay of the multi-frequency light curves.

\subsection{Scenarios A-B: single-zone expansion models}\label{sub_sec_AB}

\paragraph{Scenario A: expansion of the initial electron/positron} population.
We first assume that the electron/positron population produced during the 2017 neutrino/$\gamma$-ray flare is the only source of radiating particles, and we evolve it under adiabatic and radiative losses during blob expansion. We explored a broad parameter space in expansion speed and magnetic-field decay profiles, $V_{\rm exp}$ and $m_{\rm b}$, and computed 1.2-22\, GHz model light curves (see Appendix~\ref{App:AB}).

Across the explored parameter space, the models fall into two generic regimes (illustrated with representative examples in Appendix~\ref{App:AB}, in Fig.~\ref{fig:app_AB}). If the magnetic field decays slowly (small $m_{\rm b}$), it remains strong during expansion and the radiative cooling is efficient, causing the GHz emission to peak too early and fade too quickly compared to the observed $\sim$2.5\,yr delay, since electrons/positrons cool to lower energies. Conversely, if the magnetic field decays fast (large $m_{\rm b}$), the electrons/positrons cool more slowly but the synchrotron emissivity drops rapidly ($P_{\rm syn}\propto B^2$), leading to fluxes that remain below the observed RATAN-600 levels. We therefore do not find any combination of $(V_{\rm exp},m_{\rm b})$ that simultaneously reproduces the observed delay, Jy-level peak fluxes, and the overall flare morphology. The emission from the sole electrons/positrons that remain in the expanding blob is insufficient, and thus, further particle injection should be introduced.

\paragraph{Scenario B: continuous injection with escape.}
Next, we allowed for continuous injection of fresh electrons/positrons into the expanding blob with injection profile $L'_{\rm e}\propto R'^{-m_{\rm l}}$, and included particle escape on the comoving light-crossing time, $t'_{\rm cr}=R'/c$. Continuous injection can sustain radio-emitting electrons/positrons at later times and can be tuned to reproduce the peak time and amplitude in individual bands for suitable parameters (see Appendix~\ref{App:AB}). However, a detailed comparison to the multi-frequency RATAN-600 light curves shows the following two discrepancies with the observed radio data. First, the common year-scale radio outburst envelope of the peaks across the GHz bands is difficult to reproduce in a flare driven by the evolution of SSA, which generally predicts that higher frequencies peak earlier than lower ones as the turnover shifts through the radio band \citep{2022A&A...657A..20B} (see also Appendix~\ref{app_sec_kin}). Moreover, the radio spectral break remains somewhat constant throughout the radio outburst, and the pre-break slope is softer than the expected value from SSA (i.e., 2.5; \cite{1986rpa..book.....R}), as shown in Appendix \ref{radio_slope}.
Second, the observed radio decline in all bands at $\sim 2021$ (see Fig.~\ref{fig:sigma_vs_p}) is faster than the model prediction (see solid lines in Fig.~\ref{fig:app_AB}), which is dictated by the light-crossing time of the emitting region at the radio peak ($R'/c$), and is also much shorter than the expectation from adiabatic timescale alone ($t'_{\rm ad}= R'/V_{\rm exp}>R'/c$). Matching the rapid decay would require introducing an additional ad hoc change in the microphysics (e.g., an abrupt shutdown of injection and/or escape on timescales much faster than $R'/c$), which is not naturally motivated within this framework.

\subsection{Scenario C: Optically thin re-dissipation and blob deceleration}\label{sub_sec_C}

Motivated by the negative results of scenarios A \& B, we consider a more complex scenario as a natural extension, in which the delayed radio flare is produced by a renewed particle energization episode in the optically thin regime. We assume that the initial energy dissipation event associated with the 2017 neutrino/$\gamma$-ray flare launches a propagating dynamical disturbance in the jet which is advected outward, expands, and eventually decelerates. By ``disturbance'', we mean a localized perturbation of the flow, whose internal energy,  particle pressure, and/or magnetic pressure differs from that of the surrounding quiescent jet plasma. When this disturbance reaches parsec scales, it triggers renewed particle acceleration (e.g., at a standing shock or reconfinement shock) and channels energy into the local lepton reservoir, which may include pre-existing low-energy leptons in the jet plasma and fossil non-thermal particles from previous activity episodes, producing the delayed radio flare. We do not distinguish between these two seed populations in the modeling. The key point is that the disturbance launched during the 2017 neutrino flare triggers renewed dissipation and particle acceleration at larger distances. We discuss physical realizations of the re-dissipation in Sec~\ref{sec:disc}.

To quantify this scenario with {\tt LeHaMoC}, we proceed in three steps. (i) We first construct the post-flare particle distribution by injecting the flare-state broken power-law electrons/positrons (Table~\ref{tab:blob_params}) into a stationary zone with fixed $(R'_0,B'_0,\delta_0)$ and evolving the kinetic equations until a steady-state distribution is reached at $t'=10\,R'_0/c$. This step provides an initial condition representative of the particle population at the end of the 2017 flare. (ii) We then follow an expansion phase without additional injection or escape (as in Scenario~A), during which the particle distribution evolves under adiabatic and radiative losses. This phase is required for the SSA turnover to drop below the RATAN-600 bands so that the subsequent flare is produced in the optically thin regime. (iii) At a comoving time $t'_{\rm reacc}$, we introduce a phenomenological injection term representing delayed downstream re-dissipation. For simplicity, the injected particles are assumed to have the same broken power-law injection shape as in the 2017 flare (i.e., the same $\gamma_{\rm e,min}$, $\gamma_{\rm e,b}$, $\gamma_{\rm e,max}$, $s_{\rm e,1}$, and $s_{\rm e,2}$ as in Table~\ref{tab:blob_params}), with the injected comoving luminosity given by
\begin{equation}
    L'_{\rm inj}(R')=L'_{\rm reacc}\left(\frac{R(t'_{\rm reacc})}{R(t')} \right)^{m_{\rm l}}, \qquad t'\geq t'_{\rm reacc}
\end{equation}
so that $L'_{\rm inj}(t'_{\rm reacc})\equiv L'_{\rm reacc} $ by construction. 

We fit the multi-frequency RATAN-600 lightcurves with an MCMC procedure (Appendix~\ref{app_sec_mcmc}), restricting the fit to the time interval 2017.8-2021 before the onset of the steep post-peak decay. The free parameters are the expansion speed $V_{\rm exp}$, the onset time of re-dissipation $t'_{\rm reacc}$, the injection index $m_{\rm l}$, and the magnetic-field decay index $m_{\rm b}$ defined through $B'(R')\propto R'^{-m_{\rm b}}$. The prior on $V_{\rm exp}$ is restricted to the range inferred from the VLBI 15\,GHz core-size evolution (Appendix~\ref{Appb}). We adopt uniform priors $m_{\rm b}\in[0,2]$, spanning the standard range between toroidal-like and poloidal-like MHD scalings in a relativistic jet \citep{2011PhRvE..83a6302L}, and $m_{\rm l}\in[-1,2]$ where positive values describe declining dissipation during expansion and negative values value allow for delayed energy dissipation at a downstream site.

We assume that the emitting disturbance fills (order-unity fraction of) the local jet cross section (see sketch in Fig.~\ref{fig:sigma_vs_p}). In addition to the parameter bounds, we impose a conservative energetics prior during the re-dissipation phase to avoid strongly particle-dominated solutions, requiring that the injected electrons/positrons energy density is at most comparable to the local magnetic energy density. For an emitting region of size $R'$, this translates into a constraint of the form, 
\begin{equation}
    L'_{\rm inj}(R') \lesssim \pi R'^2 c\,U'_B(R'),
\label{eq:inj_budget}
\end{equation}
assuming that the injected non-thermal electron/positron energy is accumulated over a dynamical time of order $t'_{\rm dyn}\sim R'/c$.

\begin{figure}
\centering
\includegraphics[width=\columnwidth]{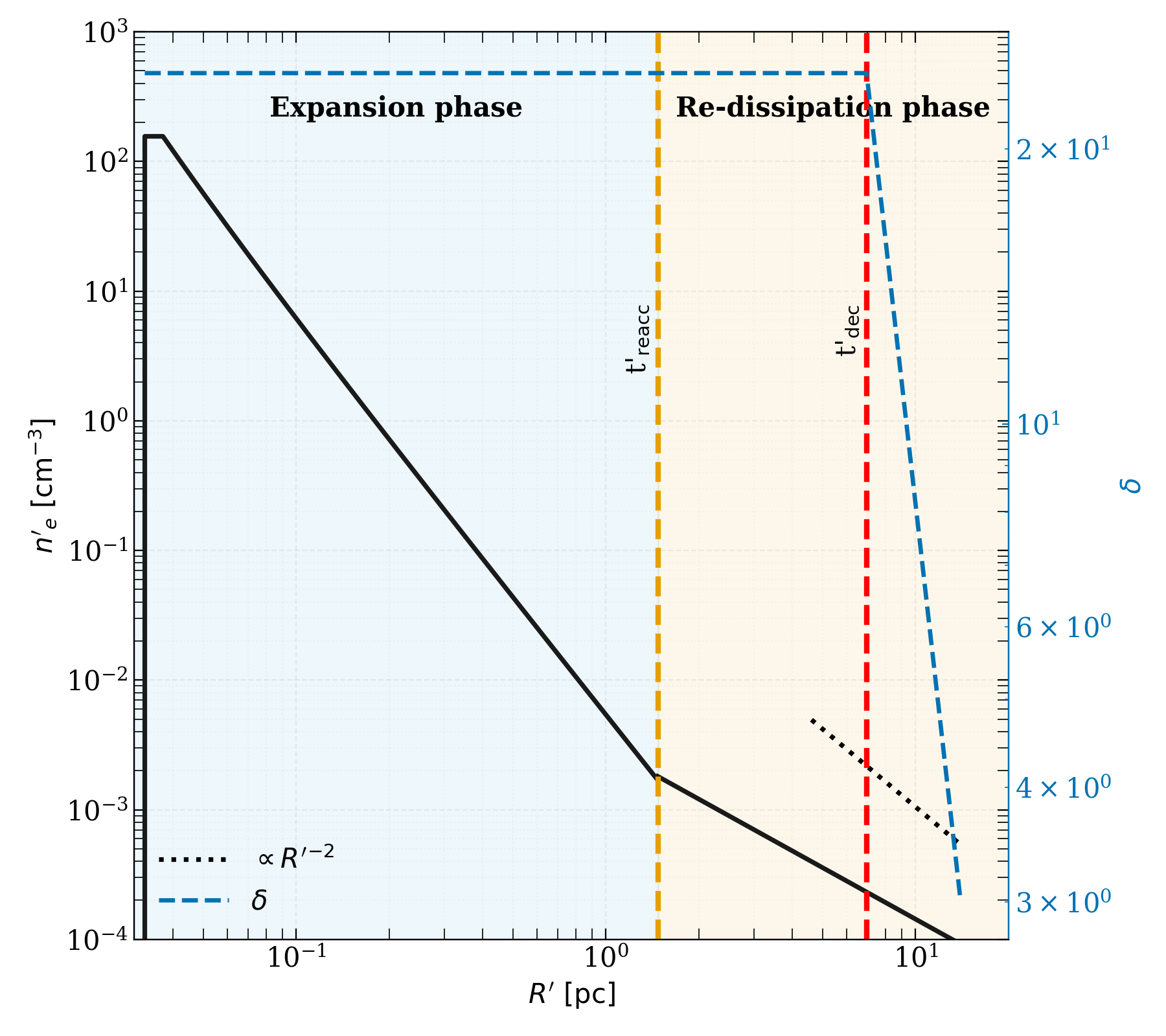} 
    \caption{Profiles of the comoving non-thermal electron/positron number density 
$n'_{\rm e}$ (left axis, solid black) together with the Doppler factor $\delta(R')$ (right axis, blue dashed) as a function of the radius of the emitting region $R'$. The dotted black line shows the asymptotic $R'^{-2}$ scaling at large radii. Vertical dashed lines mark the transition to the reacceleration phase at $t'_{\rm reacc}$ and the deceleration time $t'_{\rm dec}$. Shaded regions indicate the expansion and re-dissipation phases.}
\label{fig:n_e_delta}
\end{figure}

The normalization $L'_{\rm reacc}$ is not treated as an independent free parameter. For each trial set ($V_{\rm exp},\ t'_{\rm reacc},\  m_{\rm l}$, and $m_{\rm b}$) we fix $L'_{\rm reacc}$ by requiring continuity of the particle number density at the transition from the expansion phase to the re-dissipation, i.e. $n'_{\rm e}(t'^-_{\rm reacc})=n'_{\rm e}(t'^+_{\rm reacc})$, thereby avoiding artificial density jumps at $t'_{\rm reacc}$. The corresponding evolution of the non-thermal electron/positron density is shown with the black solid line in Fig.~\ref{fig:n_e_delta}. The MCMC fit to the RATAN-600 light curves in the interval 2017.8-2021 yields posterior median values of $V_{\rm exp}=0.28c$, $t'_{\rm reacc}=172.93R'_0/c$, $m_{\rm l}=-0.68$, and $m_{\rm b}=0.56$. With this parametrization, the negative $m_{\rm l}$ implies that the injected power increases with radius after the onset of re-dissipation, while the magnetic field decays.

\begin{figure}
\centering
\includegraphics[width=\columnwidth]{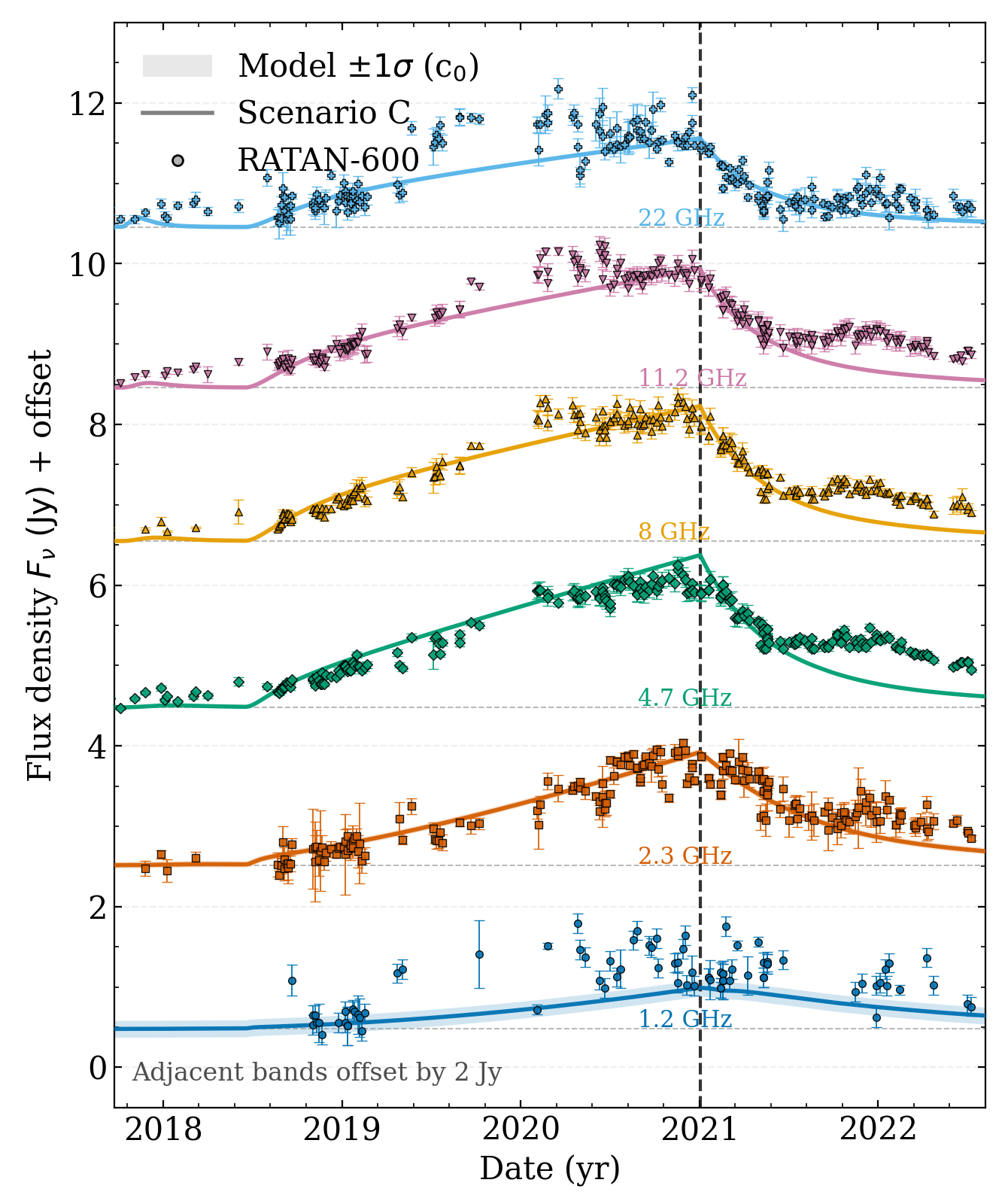}
\caption{Multi-frequency RATAN-600 radio light curves of TXS~0506+056 at 1.2–22 GHz (markers with 1$\sigma$ uncertainties), shown with vertical offsets of 2 Jy between adjacent bands for clarity. The solid curves correspond to Scenario C (re-dissipation followed by jet deceleration). The vertical dashed line marks the onset of the deceleration phase, after which the Doppler factor is assumed to decrease as a power law. For each band, the thin horizontal dashed line indicates the estimated quiescent/background level.}
\label{fig:models_c}
\end{figure}

In Fig.~\ref{fig:models_c}, we indicate for each band the quiescent/background flux level (thin dashed lines), which we estimate independently by fitting each light curve with a Gaussian flare profile plus a constant offset (see Eq.~\ref{eq:c_0_b_eq} in Appendix~\ref{app_sec_mcmc}). While this model reproduces the rise and peak of the radio-flare, a constant Doppler factor cannot account for the rapid post-peak decay observed in all radio bands (the decay timescale is much shorter than the light-crossing time at the corresponding radii). Motivated by VLBI evidence that the inner jet of TXS~0506+056 likely slows down and/or is velocity-stratified (spine–sheath) on sub-mas scales \citep{2020A&A...633L...1R}, we introduce a bulk-deceleration episode starting at $t'_{\rm dec}\simeq 750 R'_0/c$, parametrized as $\delta(R')\propto R'^{-m_{\rm d}} $. We find that a steep deceleration with $m_{\rm d}\simeq 3$ is required to reproduce the fast decay of the radio lightcurve (see blue dashed line in Fig.~\ref{fig:n_e_delta}). The 1.2 GHz data are included in the MCMC fit, but the small pre-2020 excess in this band should not be overinterpreted as a failure of Scenario C. Low-frequency blazar emission can contain slowly varying larger-scale or diffuse jet emission in addition to the compact flare component \citep[e.g.,][]{2018ApJ...869..133F, 2022A&A...663A.129N}. In Table~\ref{tab:radio_zone_derived} we list representative derived quantities at the onset of re-dissipation and at the beginning of the Doppler-factor decline. These values indicate that the radio flare is produced once the disturbance has reached parsec scales in size, and the magnetic field has decreased to $B'\sim 10^{-2}-10^{-1}\,{\rm G}$, comparable to values inferred from VLBI core-shift studies of parsec-scale blazar jets \citep{2009MNRAS.400...26O}. The density $n'_{e}$ refers only to the modeled non-thermal electron/positron population\footnote{The total plasma $\beta$ and true magnetization depend on the unmodeled cold/thermal particles of the jet.}, while $\gamma_{\rm syn}$ corresponds to the Lorentz factors of electrons/positrons that emit in the 1.2-22GHz band.

\begin{table}
\centering
\caption{Representative derived physical conditions of the radio-emitting region in Scenario C. Reported values are evaluated using the median values of the posterior distributions of fitted parameters listed in Table~1}
\label{tab:radio_zone_derived}
\begin{tabular}{lcc}
\hline
Quantity & At $t'_{\rm reacc}$ & At $t'_{\rm dec}$ \\
\hline

$R'$ [pc] & $1.6$ & $6.8$ \\
$B'$ [G] & $4.5\times10^{-2}$ & $2.0\times10^{-2}$ \\
$n'_{\rm e}$ [cm$^{-3}$] & $1.9\times 10^{-3}$ & $2.4\times 10^{-4}$\\
$\delta$ & $24.2$ & starts decreasing \\
$\gamma_{\rm syn}(1.2$--$22\,{\rm GHz})$ & $\sim 20$--$85$ & $\sim 30$--$127$ \\
\hline
\end{tabular}
\end{table}

\section{Discussion}\label{sec:disc}

TXS~0506+056 and PKS~0446+11 illustrate that neutrino-associated blazar activity can differ in $\gamma$/neutrino - radio timing patterns, from nearly contemporaneous multi-band flares \citep{2025arXiv251107535K}, which is a typical $\gamma$–radio delay statistics \citep{2015MNRAS.452.1280R} in blazars, to multi-year delayed radio outbursts. Modeling of PKS 0446+11 suggests an extremely small viewing angle ($<1^\circ$) and a rapid increase in Doppler factor \citep{2025arXiv251107535K}, whereas the multi-year delay in TXS~0506+056 points to a different jet geometry or propagation, where the disturbance must traverse a substantial distance and overcome strong SSA before becoming visible at GHz frequencies (e.g., upon reaching a parsec-scale radio “core”). This is consistent with TXS~0506+056 being viewed at a more moderate angle of several degrees \citep{2019MNRAS.483L..42K}. While rapid multi-wavelength campaigns remain essential \citep{2017APh....92...30A}, longer-term radio and VLBI monitoring are valuable for testing whether the same disturbance later produces parsec-scale flux, structural, or polarization changes downstream.

The detection of a high-energy neutrino from TXS~0506+056 implies the presence of ultra-relativistic protons (hadronic population) in the jet. These protons can interact with ambient photons to produce charged pions and hence neutrinos through p$\gamma$ interactions. In these leptohadronic models \citep{Keivani2018ApJ,2018ApJ...863L..10A,2019NatAs...3...88G,2019MNRAS.483L..12C,2022ApJ...927..197A}, the electromagnetic emission is driven by electron/positron emission, which is assumed to be present in the same emitting region. Proton synchrotron radiation is negligible at GHz frequencies; if we include in our model a pre-accelerated proton distribution with the same dynamical evolution as in Scenario~C and adopt the LMBB2b proton parameters (Table~7 of \citealp{Keivani2018ApJ}), the proton-synchrotron component peaks at $\nu F_\nu \sim 10^{15}$–$10^{16}$~Hz during the optically thin re-dissipation phase, contributing negligibly to the radio band. We also find that the number density of secondary pairs remains small compared to the primary electrons/positrons, $n'_{\rm sec}/n'_{\rm prim}\sim10^{-5}$, during the flare. Since photohadronic and Bethe-Heitler pair injection is most efficient in the compact inner zone, where the target photon density is highest, the secondary pair contribution is expected to be even less important once the disturbance has propagated further downstream.

A delayed GHz radio-flare can be understood as the downstream “afterglow” of a disturbance that is initially optically thick and at later times becomes optically thin. In expanding-source, the SSA turnover drifts to lower frequencies as the emission region expands and the magnetic field decreases, producing frequency-dependent lags when the flare is opacity-driven \citep{1966Natur.211.1131V, 2015A&A...580A..94F, 2022A&A...657A..20B, 2022A&A...658A.173T,  2023A&A...669A.151Z}. In TXS~0506+056, the observed radio evolution of the RATAN-600 peaks across 1.2–22 GHz is against a flare driven by the slow passage of the SSA turnover through the observing bands, and instead motivates (i) an expansion phase long enough for the turnover to drop below the GHz regime, followed by (ii) renewed dissipation at parsec scales (e.g. at a standing/reconfinement shock), which naturally leads to an optically thin radio outburst with no inter-band delays. A local episode of particle acceleration in a fraction of the jet plasma could also produce a radio-flare without requiring the original compact emitting region itself to propagate downstream. The role of the propagating disturbance is instead to provide the causal link to the 2017 neutrino/$\gamma$-ray event to a downstream dissipation site where the delayed radio-flare is produced. Within this interpretation, the expansion speed sets how quickly the source reaches the optically thin regime, while the magnetic-field decay controls both the SSA evolution and the available synchrotron power. Our fit favors a mildly relativistic expansion speed ($V_{\rm exp} \sim 0.3 c$) together with an effective field decline ($m_{\rm b} \sim 0.6$), and the rapid post-peak decay further points to Doppler-factor evolution, such as jet deceleration or velocity stratification \citep{2000A&A...361..850T,2003ApJ...589L...5G}.

Relativistic hydrodynamic (HD) and MHD works provide a physical context for our scenario in which the disturbance launched during the neutrino $\gamma$-ray episode produces a delayed flare only after propagating to larger distances. In our case, the observed delay of $\Delta t_{\rm obs}\simeq 2.5$\,yr corresponds (for constant $\delta_0$ and ballistic motion) to a deprojected propagation distance $\Delta z = \beta c\,\Delta t_{\rm obs}/[(1+z_{\rm cos})(1-\beta\cos\theta)]$, where $\theta$ is the jet viewing angle, corresponding to $\Delta z \simeq 0.57\,{\rm pc}\times (\delta_0\Gamma_0)$ for $z_{\rm cos}=0.3365$,
i.e. tens to a few hundreds of pc, for the parameters used above. Simulations of jets show that moving compressions/shocks or “blobs” can transport energy downstream without any radiative signature, and then dissipate efficiently when they encounter pre-existing jet structures such as standing recollimation shocks, shear layers, or other quasi-stationary features, producing a secondary flare via shock–shock interaction and renewed particle energization (e.g., \citealp{2016A&A...588A.101F,2021A&A...647A..77F}). \cite{2024ApJ...976..144D} found that shocks in a relativistic MHD jet can sustain particle acceleration over pc scales. In their simulations with diffusive shock acceleration, electrons were accelerated in numerous shock fronts, producing bright knots and moving features that flare as they evolve.  In this picture, time delays between the inner-jet high-energy flare and a later GHz outburst arise naturally from the propagation time to a downstream dissipation zone and from changes in beaming if the flow decelerates or becomes stratified.

\section{Conclusions}

We investigated whether the multiyear delay between IceCube-170922A and the subsequent GHz radio flare in TXS~0506+056 can be explained by the post-2017 evolution of the compact high-energy region associated with the neutrino event. In particular, we tested whether this region, initialized with leptohadronic flare parameters inferred for the 2017 event, can reproduce the 1.2-22 GHz RATAN-600 light curves as it expands downstream and becomes transparent to SSA.

We find that a direct adiabatic expansion of the primary electrons/positrons related to the 2017 neutrino/$\gamma$-ray flare (Scenario A) is excluded. No combination of expansion speed and magnetic-field evolution reproduces the observed 2-3 yr delay, Jy-level peak fluxes, and flare morphology. Allowing for continuous particle injection (Scenario B) improves the results but still predicts stronger frequency-dependent peak offsets than observed and fails to explain the rapid post-peak decay without fine-tuned changes in the microphysics.

A more plausible interpretation is that the disturbance responsible for the 2017 neutrino/$\gamma$-ray flare exhibits re-dissipation of energy downstream, after the source has become optically thin at GHz frequencies. In this scenario (Scenario C), renewed particle acceleration reproduces the multi-frequency radio evolution. The rapid decline of the 2020 flare requires a decrease of the Doppler factor once the emitting region has reached the large sizes implied by the delayed radio outburst, since a constant-Doppler solution would decay on a much longer timescale. This is consistent with a decelerating and/or velocity-stratified jet.

Within a one-zone time-dependent framework, the delayed radio-flare in TXS~0506+056 is unlikely to be a direct radiative echo of the neutrino-production region becoming transparent. Instead, the delayed GHz emission primarily traces downstream dissipation and beaming evolution of the propagating disturbance. Because the radio emission is dominated by leptonic synchrotron emission and is largely insensitive to the proton population relevant for neutrino production, a delayed radio-flare is not necessarily linked to the proton distribution characteristics, but rather a diagnostic of how and where the jet reprocesses energy at parsec scales. This informs future modeling of neutrino–blazar associations by indicating that radio-flares appearing years after a neutrino alert may trace downstream dissipation and Doppler-factor evolution, rather than direct radio unveiling of the neutrino zone. It also motivates long-term radio monitoring (months–years) and high-resolution VLBI follow-up after neutrino alerts to capture delayed flares and associated structural changes.

\begin{acknowledgements}
We are grateful to Apostolos Mastichiadis, Lea Marcotulli, Yuri Kovalev, Xavier Rodrigues, and Damiano F.G. Fiorillo for the useful comments and insightful conversations. The research project has been funded by the ``Program for the Promotion of Exchanges and Scientific Collaboration between Greece and Germany IKYDA--DAAD'' 2024 (IKY project ID 309; DAAD project ID: 57729829). CY acknowledges the support of the Institut Interuniversitaire
des Sciences Nucl\'eaires (IISN) Grant No. 4.4503.15.
\end{acknowledgements}

\appendix
\raggedbottom
\section{ Scenarios A and B}\label{App:AB}

\subsection{Scenario A}\label{App:ScenA}
\begin{figure*}
    \centering {\includegraphics [width=0.85\textwidth]{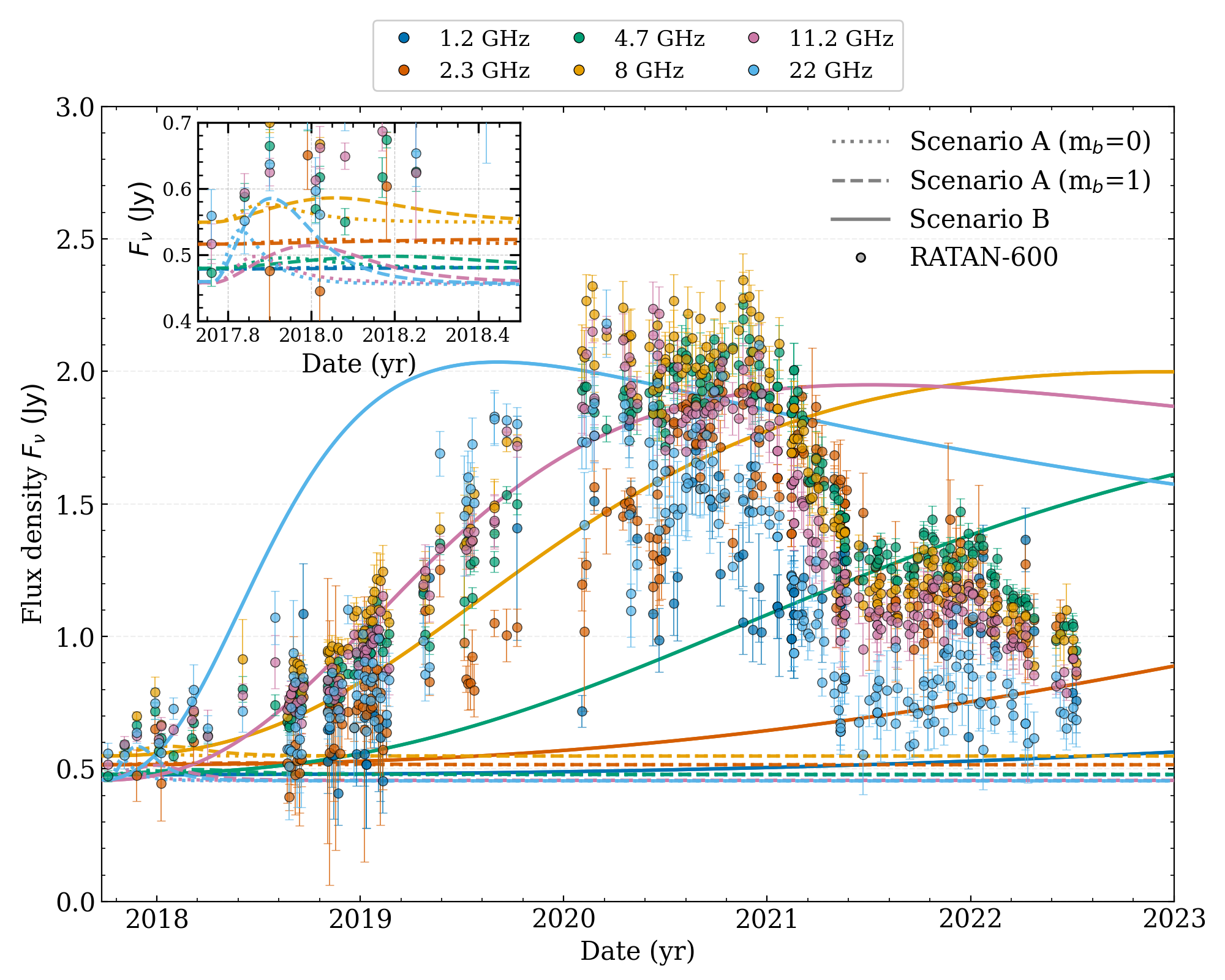}}
    \caption{Representative synthetic multi-frequency (1.2-22\,GHz) light curves for Scenario A (dotted and dashed lines) and B (solid line), same as Fig.~\ref{fig:sigma_vs_p}.}
    \label{fig:app_AB}
\end{figure*}

In Scenario A, the emitting region is populated by relativistic electrons/positrons that were initially injected during the neutrino/ $\gamma$-ray flare without any further injection or escape during the expansion period. The blob initially has radius $R'_0$, magnetic field $B'_0$, and Doppler factor $\delta_0$, and is occupied with a broken power-law distribution of primary electrons/positrons (see Table~\ref{tab:blob_params}). We evolve the electron/positron distribution until it reaches steady-state ($t'=10 R_0'/c \Rightarrow t_{\rm obs}=(1+z)t'/\delta_0\simeq 0.06~\rm yr$, with $z=0.3365$ being the redshift of TXS~0506+056 \cite{2018ApJ...854L..32P}) while keeping the radius and the magnetic field strength equal to their initial values. We then let it expand at a speed $V_{\rm exp}$ and evolve the particle distribution under adiabatic and radiative losses. We assume that the magnetic field in this case scales as $B'(R')\propto R'^{-m_{\rm b}}$. We explored a broad range of these parameters $(V_{\rm exp} \in [0.01, 0.9]c$ and $m_{\rm b} \in [0.,2.]$), and for each model computed synthetic 1.2–22. GHz light curves. The parameter exploration reveals two generic behaviors, illustrated in Fig.~\ref{fig:app_AB} with representative model realizations in which we assume that the expansion velocity is $V_{\rm exp}=0.2 c$\footnote{We adopt a fixed $V_{\rm exp}$ in Fig.~\ref{fig:app_AB} to highlight the dependence on $m_{\rm b}$. Varying $V_{\rm exp}$ primarily shifts the model light curves in observer time (and changes the normalization modestly), but does not remove the two generic behaviors described below. Lower expansion velocities shift the radio peak to later times but reduce the flux density since they correspond to more efficient escape of photons.}:

(i) Weak field decay (small $m_{\rm b}$): the magnetic field remains almost constant or decreases mildly during expansion, and synchrotron cooling is efficient ($t'_{\rm syn}\propto B'^{-2}$). The electron/positron distribution cools rapidly, and the GHz synchrotron light curves peak early (soon after the $\gamma$-ray flare) and fade on timescales much shorter than the observed delay.

(ii) Steep field decay (large $m_{\rm b}$): radiative cooling becomes less effective, but the synchrotron emissivity drops quickly because $P_{\rm syn}\propto B^{\prime 2}$. As a result, the model fluxes remain below the observed RATAN-600 levels.

\subsection{Scenario B}\label{App:ScenB}
In Scenario B, we include continuous particle injection, and we allow for the escape of electrons/positrons during the expansion phase. Initially we evolve the particle distribution until a steady state is reached, as in Section~\ref{App:ScenA}, but during the expansion phase, we allow for further injection of electrons into the blob. The comoving electron/positron injection power is parameterised as $L_{\rm e}'(R')\propto R'^{-m_{\rm l}}$, and we include a simple prescription for particle escape equal to the comoving light-crossing time $t'_{\rm cr}=R'/c$. The expansion speed $V_{\rm exp}$ and magnetic-field decay index $m_{\rm b}$ are again treated as free parameters. With continuous injection, the model can maintain a significant population of radio-emitting electrons/positrons at late times. We find that for magnetic field decay with $m_{\rm b}\simeq 1$ and slowly decreasing injection power $m_l\simeq 0.1$, it is possible to reproduce the properties of the flare in individual bands for suitable choices of $V_{\rm exp}$ ($V_{\rm exp}\simeq 0.05c$). The 11–22 GHz light curves reach Jy-level maxima after 2-3 yr (see solid curve in Fig.~\ref{fig:app_AB}), similar to the RATAN-600 data, and exhibit a roughly symmetric rise and fall\footnote{We adopt $V_{\rm exp}\simeq 0.05c$ here because it provides a near-optimal match to the observed delay and Jy-level peaks in the high-frequency RATAN-600 bands within this single-zone framework; varying $V_{\rm exp}$ primarily shifts the model light curves in time but does not remove the inconsistencies discussed below.}. 
However, when we compare the model in detail with the multi-frequency RATAN-600 light curves, two inconsistencies appear. All radio light curves peak within $\sim$ 0.5 years of each other. In a scenario where the flare is driven by $\nu'_{\rm ssa}$, one expects higher frequencies to peak earlier than lower ones, as the SSA turnover/break gradually moves. Analytically one can show that the SSA frequency in the comoving system evolves as,
\begin{equation}
\nu'_{\rm ssa}\propto R'^{-\left(1+m_{\rm b}\frac{s_{\rm e,1}+1}{2}+m_{\rm l}\right)\frac{2}{s_{\rm e,1}+4}},
\end{equation}
assuming that the electrons/positrons before the spectral break $\gamma_{\rm br}$ are responsible for SSA (for the derivation of the formula, see Appendix~\ref{app_sec_kin}). Our model can explain such year-scale multi-frequency outburst envelope if $\nu'_{\rm ssa}$ drops extremely rapidly across the GHz range. Such a drop would mean that either the magnetic field or the injected electrons/positrons luminosity should decrease fast as the blob expands (large values of $m_{\rm b}$ and $m_{\rm l}$). This would lead to an inconsistency because of the low flux levels in such a physical scenario. The second inconsistency in this physical scenario is the decay timescale. The characteristic decay timescales in the RATAN-600 bands are comparable to, or shorter than, the light-crossing time of the emitting region at peak, $R'/c$, and much shorter than the expectation from adiabatic cooling alone. Within the framework of a single expanding blob with a constant Doppler factor, it is difficult to obtain such a rapid decay without invoking unphysical changes in the microphysics (e.g. abrupt cessation of injection and extremely fast escape).

\section{Expansion velocity of the emitting region}\label{Appb}

\begin{figure*}[ht!]
    \centering {\includegraphics [width=0.8\textwidth]{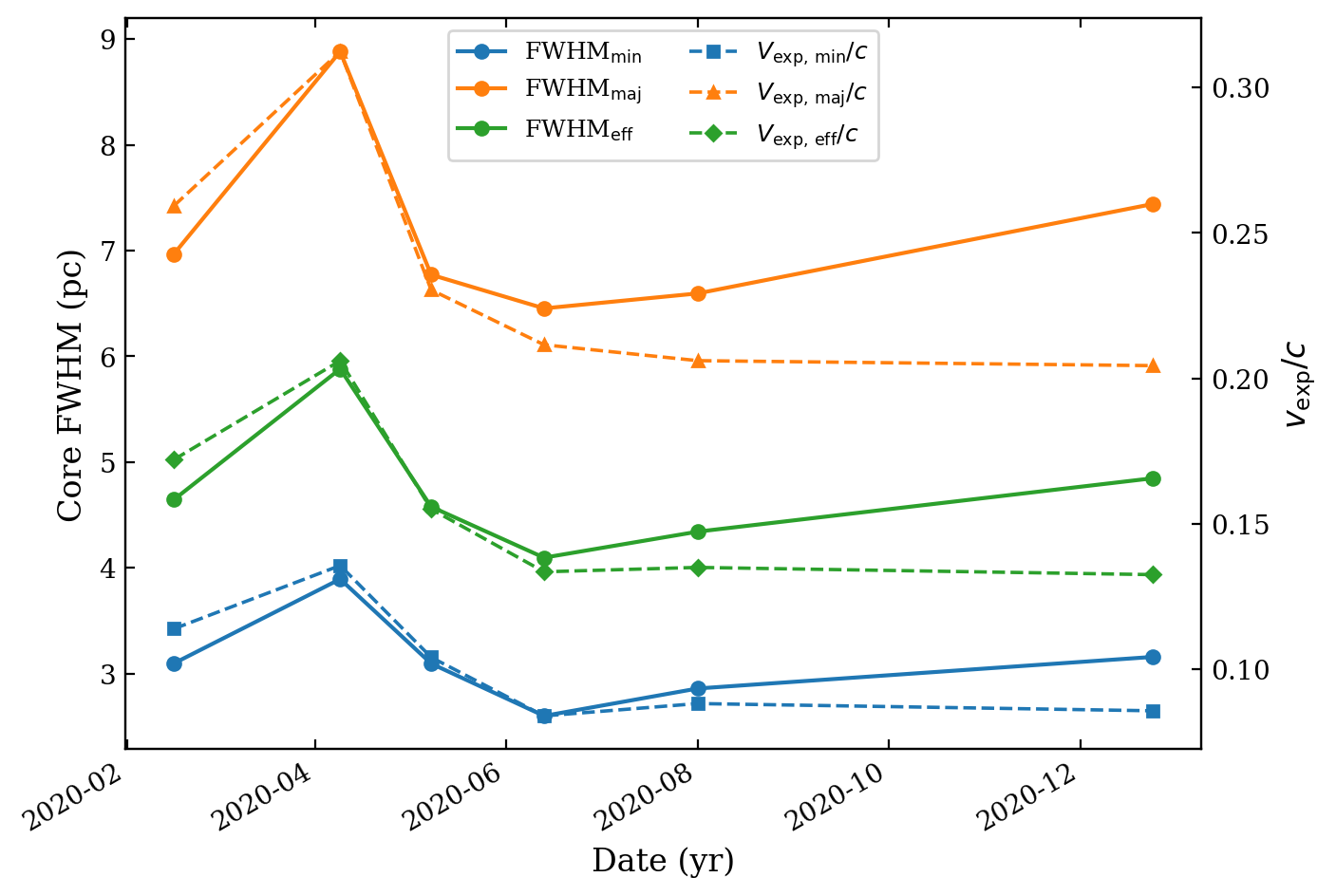}}
     \caption{Evolution of the VLBI core size and inferred expansion velocity of TXS~0506+056 at 15~GHz in 2020. Solid curves show ${\rm FWHM}_{\rm min}$, ${\rm FWHM}_{\rm maj}$ and ${\rm FWHM}_{\rm eff}$ in parsecs (left axis), while dashed curves show the corresponding expansion velocities in units of $c$ (right axis).}
      \label{fig:Blob_size_Exp_Vel}
\end{figure*}

To constrain the expansion velocity of the emitting region, we use the publicly available 15 GHz VLBA MOJAVE\footnote{\url{https://www.cv.nrao.edu/MOJAVE/}} database images of TXS~0506+056 \citep{2018ApJS..234...12L}, and specifically the naturally weighted Stokes I CLEAN maps for the year 2020 from the MOJAVE source page, during which the source exhibits a pronounced brightening of the radio core. We select all available epochs in 2020 and quantify the structural evolution of the core at the peak of the radio-flare. For each epoch, we isolate a small region around the VLBI core and fit a single two-dimensional elliptical Gaussian plus a constant background directly to the clean image, using a non-linear least-squares minimization. From this fit we obtain the full width at half maximum (FWHM) along the major and minor axes, $ {\rm FWHM}_{\rm maj}$ and ${\rm FWHM}_{\rm min}$, as well as the effective FWHM defined as.
\begin{equation}
{\rm FWHM}_{\rm eff} = \sqrt{{\rm FWHM}_{\rm maj} \times {\rm FWHM}_{\rm min}}.
\end{equation}
The temporal evolution of these three quantities is shown in Fig.~\ref{fig:Blob_size_Exp_Vel}. TXS~0506+056 is located at redshift $z_{\rm cos} = 0.3365$. Adopting a Planck–like cosmology, the MOJAVE team and \cite{2020A&A...633L...1R} quote a linear scale of 1~mas $= 4.78$~pc at the distance of TXS~0506+056, which we use to convert angular sizes into sky-plane (transverse) physical sizes. Interpreting the brightened 15 GHz core during the 2020 outburst maximum as being dominated by the flare component, we use the fitted Gaussian size as a proxy for the transverse extent of the dominant emitting region (blob) at that epoch\footnote{In general, the VLBI “core” at a fixed frequency is often identified with the $\tau_\nu\simeq 1$ surface of a partially self-absorbed jet \citep{1979ApJ...232...34B}}. Our working assumption here is that near the flare maximum, the outburst component dominates the core emission. We convert each of these angular sizes into a projected physical radius on the plane of the sky using that,
\begin{equation}
R_{\rm proj}({\rm axis}) \equiv \frac{1}{2} {\rm FWHM}_{\rm axis}\times 4.78~{\rm pc~mas^{-1}},
\end{equation}
where $\rm FWHM_{\rm axis} \in \{FWHM_{\rm maj},\ FWHM_{\rm min},\ FWHM_{\rm eff}\}$. The resulting radii are shown in Fig.~\ref{fig:Blob_size_Exp_Vel}, providing a range of plausible transverse sizes of the emitting region during the peak of the radio-flare. Because the blob is assumed to be approximately spherical and the fitted FWHM values describe the transverse extent of the core on the sky, these radii are not deprojected by the jet viewing angle. In the modelling we therefore adopt $R'_{\rm rf}\simeq R_{\rm proj}$, using the spread between the minor-, major-, and effective-axis estimates as an indication of the systematic uncertainty on the characteristic size.

For each epoch we estimate the off-source image noise level following \cite{2017MNRAS.468.4992P}, as the minimum of the rms measured in four corner quadrants of the map (each quadrant spanning 1/16 of the image size). In our 2020 maps the resulting noise levels are of order $\sigma_{\rm rms}\sim(0.1$-$0.6)$\,mJy\,beam$^{-1}$, comparable to typical MOJAVE 15\,GHz images, and the bright VLBI core is detected signal-to-noise estimate, SNR (${\rm peak}/\sigma_{\rm rms}\sim10^{3}$-$10^{4}$ ). With this SNR we (i) deconvolve the fitted Gaussian FWHM from the restoring beam (assuming Gaussian shapes) and (ii) verify that the intrinsic sizes exceed the minimum resolvable size given by the resolution-limit formalism of \citep{2005astro.ph..3225L}.

To estimate the expansion speed, we compare this radio size to the characteristic size inferred from X-ray variability modelling at an earlier epoch. If the blob expands from radius $R'_0$ to $R'_{\rm rf}$ over an observed time interval $\Delta t$ between the X-ray flare and the radio core brightening, and moves with Doppler factor $\delta_0$, the corresponding expansion velocity is written as

\begin{equation}
V_{\rm exp}\simeq \frac{R'_{\rm rf}-R'_0}{\Delta t'}=\frac{(1+z)(R'_{\rm rf}-R'_0)}{\delta_0 \Delta t},
\end{equation}
corresponding to the following values $V_{\rm exp}\in (0.084, 0.31)c$ (see right y-axes of Fig.~\ref{fig:Blob_size_Exp_Vel}), using the Doppler factor presented in Table~\ref{tab:blob_params}.

\section{Kinetic solution and scaling of the SSA frequency}\label{app_sec_kin}

Here we derive the approximate dependence of the SSA turnover frequency $\nu'_{\rm ssa}$ on the radius $R'$ of an expanding, homogeneous emitting region, neglecting synchrotron and IC cooling.

\subsection*{C.1 Kinetic equation and large-$R'$ behavior}
We consider a spherical region of comoving radius $R'$, filled with a power-law electron/positron distribution at low energies. In the comoving frame, the evolution of the electron/positron distribution $N(\gamma,R')\equiv \mathrm{d}N/\mathrm{d}\gamma$ in an expanding spherical blob of radius $R'$ can be written as a function of the comoving radius $R'$ as,

\begin{equation}
\frac{\partial N(\gamma,R')}{\partial R'}
 - \frac{\gamma}{R'}\,\frac{\partial N(\gamma,R')}{\partial\gamma}
 + \frac{\beta_{\rm exp}-1}{R'}\,N(\gamma,R')
 = Q'_{\rm inj}(\gamma,R'),
\label{eq:kinetic_Rprime}
\end{equation}
where $\beta_{\rm exp}\equiv c/V_{\rm exp}$ and the second and third terms on the left-hand side describe adiabatic energy losses and escape of electrons/positrons in an expanding blob. The comoving injection term is assumed to be a power-law both in energy and in radius,
\begin{equation}
Q'_{\rm inj}(\gamma,R') = Q'_0
\left(\frac{R'_0}{R'}\right)^{m_{\rm l}} \gamma^{-p}\,
H(R'-R'_{\rm 0}),
\label{eq:Qinj_powerlaw}
\end{equation}
where $p$ is the injected electron/positron index, $m_{\rm l}$ controls the radial decay of the injected power, and $H$ is the Heaviside step function. The corresponding comoving power injected into relativistic electrons/positrons is
\begin{equation}
L'_e(R') \simeq \int \gamma m_e c^2\, Q'_{\rm inj}(\gamma,R')\,\mathrm{d}\gamma
\;\propto\;
\left(\frac{R'_0}{R'}\right)^{m_{\rm l}},
\end{equation}
in agreement with the parametrisation adopted in Sect.~3.2.

For the power-law injection in Eq.~\eqref{eq:Qinj_powerlaw} one can solve Eq.~\eqref{eq:kinetic_Rprime} by the method of characteristics. The solution that matches a power-law initial condition $N(\gamma,R'_0)=k_0\gamma^{-p}$ at $R'=R'_0$ is
\begin{equation}
N(\gamma,R)=\gamma^{-p}
\begin{cases}
k_0\left(\dfrac{R'}{R'_0}\right)^{1-p-\beta_{\rm exp}}+\dfrac{\beta_{\rm exp}}{c}\,\dfrac{Q'_0 R_0'^{1-m_{\rm l}}}{p-m_{\rm l}+\beta_{\rm exp}}
\left[\left(\dfrac{R'}{R'_0}\right)^{1-m_{\rm l}}-\left(\dfrac{R'}{R'_0}\right)^{1-p-\beta_{\rm exp}}\right],
& p-m_{\rm l}+\beta_{\rm exp}\neq 0,\\[10pt]
\left(\dfrac{R'}{R'_0}\right)^{1-p-\beta_{\rm exp}}\left[k_0+\dfrac{\beta_{\rm exp}R_0}{c}\,Q'_0\,\ln\left(\dfrac{R'}{R'_0}\right)\right],
& p-m_{\rm l}+\beta_{\rm exp}=0,
\end{cases}
\label{eq:N_solution_full_Rprime}
\end{equation}

In the regime relevant for the delayed radio-flare we are interested in radii $R'\gg R'_0$. For $\beta_{\rm exp}+p\neq m_{\rm l}$ and $R'\gg R'_0$, the first term in the square brackets of the upper branch of Eq.~\eqref{eq:N_solution_full_Rprime} dominates (for $p+\beta_{\rm exp}>m_{\rm l})$, and we obtain the asymptotic behavior,
\begin{equation}
N(\gamma,R') \;\simeq\; 
\dfrac{\beta_{\rm exp}}{c}\,\dfrac{Q'_0 R'^{1-m_{\rm l}}}{p-m_{\rm l}+\beta_{\rm exp}}\gamma^{-p},
\qquad (R'\gg R'_0,\ \beta_{\rm exp}+p\neq m_{\rm l}).
\label{eq:N_asymptotic}
\end{equation}
Thus, at large $R'$ the the differential number of electrons/positrons with respect to the Lorentz factor $\gamma$ remains a power law in energy with index $p$, and its overall normalization scales as
\begin{equation}
K_{\rm tot}(R')\propto R'^{1-m_{\rm l}}.
\end{equation}

\subsection*{C.2 Scaling of the SSA frequency}

From the asymptotic solution of the kinetic equation for large radii (Eq.~\ref{eq:N_solution_full_Rprime}), the total number of electrons/positrons per unit $\gamma$ behaves as
\begin{equation}
N(\gamma,R') \simeq C\,\gamma^{-p}\,R'^{\,1-m_{\rm l}},
\label{eq:N_asymptotic_Rprime}
\end{equation}
where $C$ is a constant and $m_{\rm l}$ is the radial index of the source term in Eq.~\eqref{eq:kinetic_Rprime}. The comoving volume is $V' = (4\pi/3)R'^3$, so the differential number density of electrons/positrons can be written as
\begin{equation}
n_e(\gamma,R') \equiv \frac{\mathrm{d}n_e}{\mathrm{d}\gamma}
= \frac{N(\gamma,R')}{V'}
= k_e(R')\,\gamma^{-p},
\end{equation}
with a radius-dependent normalization $k_e(R') \propto R'^{-(2+m_{\rm l})}$. For a homogeneous source with a power-law electron/positron distribution $n_e(\gamma)\propto k_e\,\gamma^{-p}$, the SSA coefficient in the $\delta$-function approximation can be written as \citep{1979A&A....76..306G},
\begin{equation}
\kappa_\nu^\delta
 = c_\delta(p)\,\frac{c\,r_e\,k_e}{\nu}
 \left(\frac{\nu_B}{\nu}\right)^{\frac{p+2}{2}},
\label{eq:kappa_delta}
\end{equation}
where $c_\delta(p)$ is a numerical factor that depends only on $p$,
$r_e$ is the classical electron radius, and $\nu_B=eB'/(2\pi m_e c)$ is
the cyclotron frequency. Using $\nu_B\propto B'$ and combining the
powers of $\nu$, we obtain the scaling,
\begin{equation}
\kappa_\nu^\delta \;\propto\;
k_e(R')\, B'(R')^{\frac{p+2}{2}}\,\nu^{-\frac{p+4}{2}}.
\label{eq:kappa_scaling}
\end{equation}

The SSA turnover frequency $\nu'_{\rm ssa}$ is defined by the condition that the optical depth across the source is of order unity, $\tau_{\nu'} \equiv \kappa_{\nu'}^\delta\,R' \sim 1\Rightarrow \kappa_{\nu'_{\rm ssa}}^\delta\,R' \sim 1.$ Inserting Eq.~\eqref{eq:kappa_scaling} and evaluating at
$\nu'=\nu'_{\rm ssa}$ gives $1 \;\propto\;k_e(R')\, B'(R')^{\frac{p+2}{2}}\, R'\, \nu'_{\rm ssa}{}^{-\frac{p+4}{2}},$ and hence,
\begin{equation}
\nu'_{\rm ssa}
\;\propto\;
\left[k_e(R')\, B'(R')^{\frac{p+2}{2}}\, R' \right]^{\frac{2}{p+4}}.
\label{eq:nussa_general_ke}
\end{equation}
We now use the scalings $k_e(R') \propto R'^{-(2+m_{\rm l})}$ and $B'(R') = B'_0 (R'/R'_0)^{-m_{\rm b}}$, where $m_{\rm b}$ is the radial index of the magnetic-field decay we find that $k_e(R') B'(R')^{\frac{p+2}{2}} R'\propto R'^{-(2+m_{\rm l})}\,R'^{-m_{\rm b}\frac{p+2}{2}}\,R' \propto R'^{-\left[1 + m_{\rm l} + m_{\rm b}\frac{p+2}{2}\right]}.$
Substituting this into Eq.~\eqref{eq:nussa_general_ke}, we obtain
\begin{equation}
\nu'_{\rm ssa}
\propto
R'^{-\left(1 + m_{\rm l} + m_{\rm b}\frac{p+2}{2}\right)\frac{2}{p+4}}.
\label{eq:nussa_final_m}
\end{equation}

\section{MCMC Fitting}\label{app_sec_mcmc}

We fit the multi-frequency radio light curves using Bayesian inference and explore the posterior
distribution with the affine-invariant ensemble Markov Chain Monte Carlo (MCMC) sampler
implemented in \texttt{emcee} \citep{2013PASP..125..306F}. We restrict the fit
to flux measurements within a specific time window $t \in [t_{\min}=2017.8, t_{\max}=2021]$. We set $t_{\max}$=2021 because, after this epoch, the observed light curves enter a much steeper decline that cannot be reproduced by the constant Doppler factor version of the model used in the MCMC fit. For each observing band $b$, we extract the observation times $t_i(b)$ and flux densities $y_i(b)$ and their corresponding measurement uncertainties $\sigma_i(b)$. We retain only points satisfying $t_{\min} \le t_i(b) \le t_{\max}$ and $y_i(b)$ finite.

We adopt uniform priors within the parameter ranges given in Sec.~\ref{sub_sec_C}. For each parameter vector $\boldsymbol{\theta} = \left(V_{\rm exp},\, m_{\rm l},\, t'_{\rm{reacc}},\, m_{\rm b}\right)$ we generate a model light curve using a forward model that returns (i) a model time grid $t_{\rm{model}}$, (ii) model flux predictions $F_{\rm{model}}(t_{\rm model},b)$ for each band, and (iii) quantities used for physical arguments checks (see Eq.~\ref{eq:inj_budget}).
To compare model and data at the observed epochs, we linearly interpolate the model flux in time into the observation times,
\begin{equation}
\mu_i(b;\boldsymbol{\theta}) =
\mathrm{interp}\!\left(t_i(b);\ t_{\rm{model}},\ F_{\rm{model}}(t_{\rm model},b)\right) + c_0(b),
\end{equation}
where $c_0(b)$ represents the background level in each light curve. We estimate $c_0(b)$ independently for each band by fitting a phenomenological "Gaussian flare + constant offset" model,
\begin{equation}
f_b(t) = c_0(b) + A_b \exp\!\left[-\frac{1}{2}\left(\frac{t-t_{0,b}}{s_b}\right)^2\right],
\label{eq:c_0_b_eq}
\end{equation}
using \texttt{emcee}. To account for potential additional scatter beyond the reported measurement errors, we include
a term $j_b$ added in quadrature, i.e. $\sigma_{\rm{tot,i}}^2=\sigma_i^2 + j_b^2$.
For each band we adopt the posterior median of $c_0(b)$. 

\begin{figure*}[ht!]
    \centering {\includegraphics [width=0.75\textwidth]{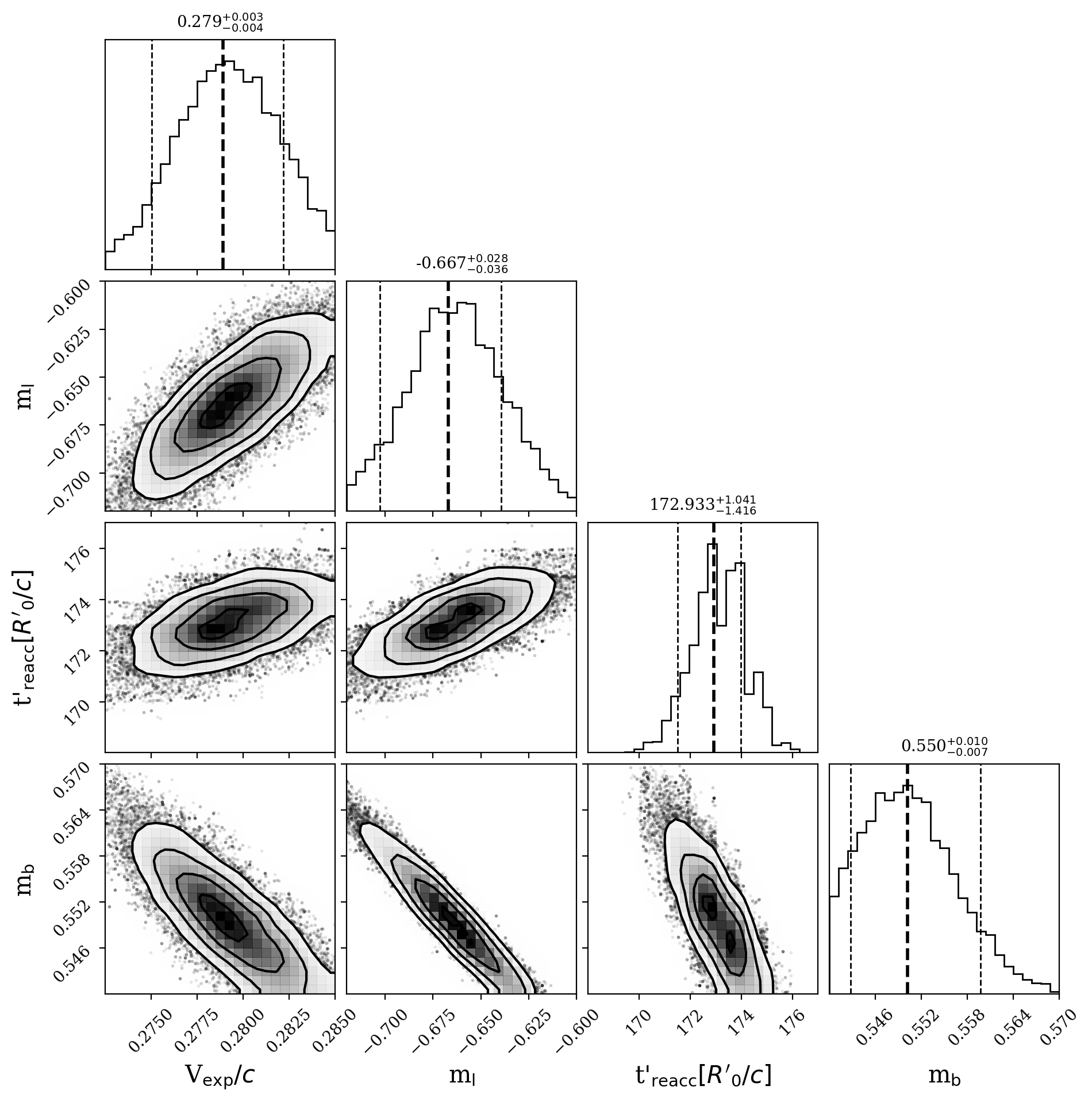}}
     \caption{Posterior distributions of the parameters described in Scenario C. Dashed lines in the one-dimensional histograms indicate the posterior median and the central $68\%$ credible interval for each parameter.}
      \label{fig:Posterior_dis}
\end{figure*}

For the radio measurements, we model their errors as independent Gaussians. The total log-likelihood is the sum over all retained data points across all bands,
\begin{equation}
\ln \mathcal{L}_{\rm{tot}}(\boldsymbol{\theta}) = -\frac{1} {2}\sum_{b}\sum_{i\in b} \left[\frac{\left(y_i(b) - \mu_i(b;\boldsymbol{\theta})\right)^2}{\sigma_{\rm tot,i}^2(b)} + \ln\!\left(2\pi\,\sigma_{\rm tot,i}^2(b)\right)\right],
\label{eq:loglike_detect}
\end{equation}
where $y_i(b)$ and $\sigma_i(b)$ are the observed flux and its uncertainty, and $\mu_i(b;\boldsymbol{\theta})$ is the model prediction after time interpolation and inclusion of the baseline $c_0(b)$. We sample the posterior with an ensemble of $32$ walkers evolved for $3000$ steps, discard the first $1500$ steps as burn-in, and use the remaining samples to compute credible intervals and parameter covariances. The posterior distributions are presented in Fig.~\ref{fig:Posterior_dis}

\section{Fitting of radio spectra }\label{radio_slope}
\begin{figure}
    \centering
    \begin{subfigure}[t]{0.45\textwidth} 
        \centering
        \includegraphics[width=\textwidth,valign=t]{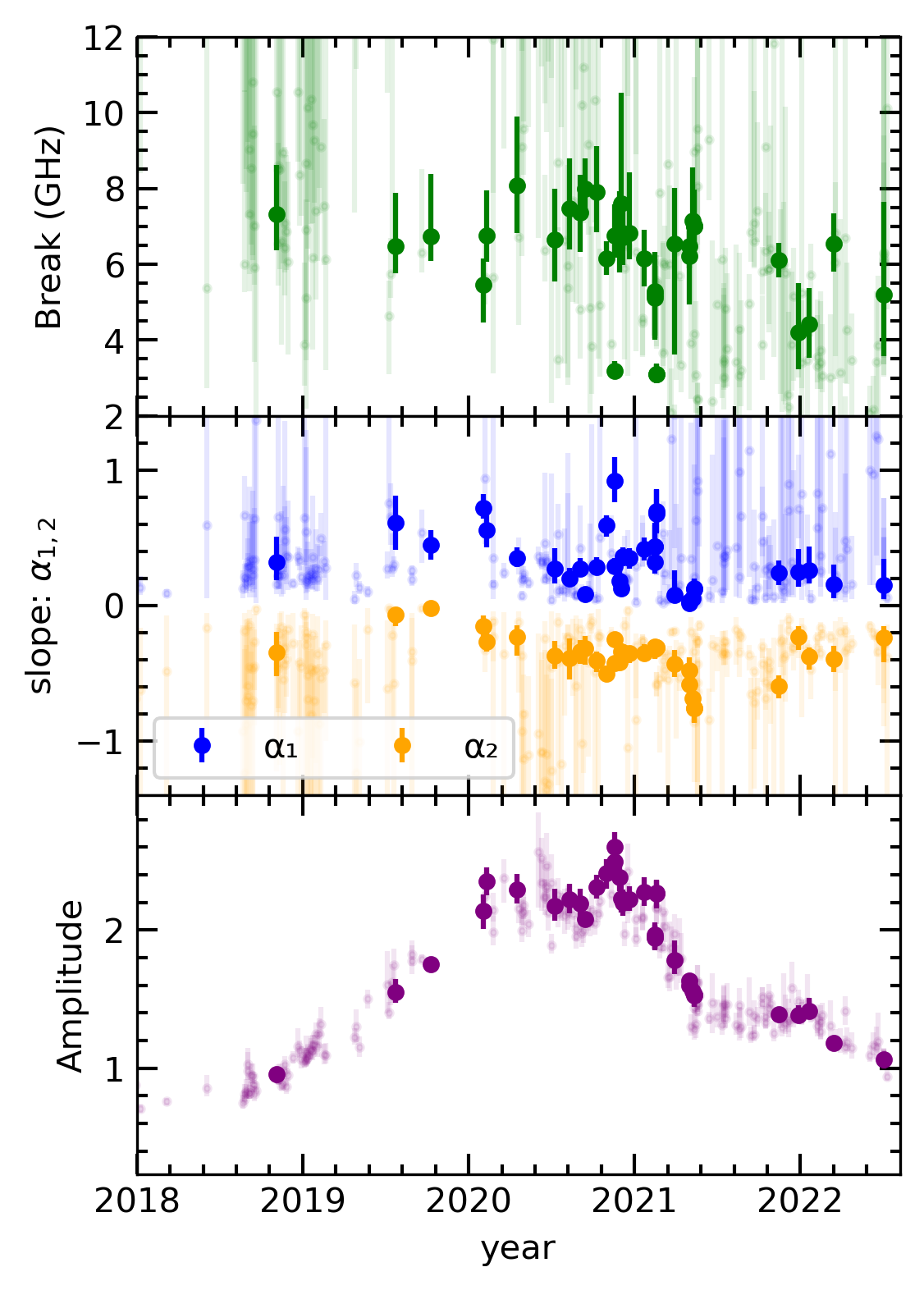}
    \end{subfigure}%
    \hfill
    \begin{subfigure}[t]{0.45\textwidth}
        \centering
        \vspace{0.3cm}
        \includegraphics[width=\textwidth,valign=t]{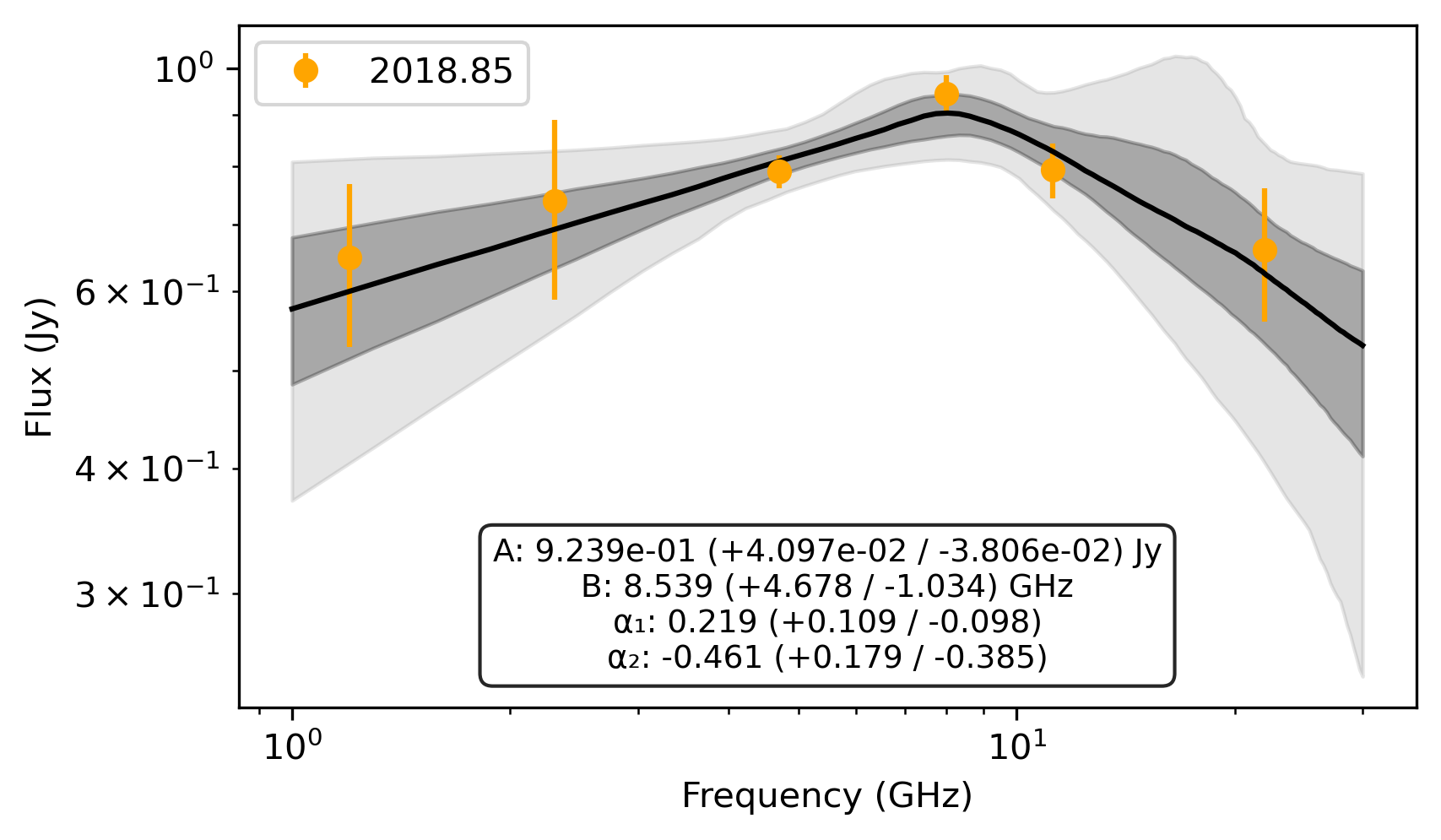}
        \label{fig:plot2}
        \vspace{2mm} 
        \includegraphics[width=\textwidth,valign=t]{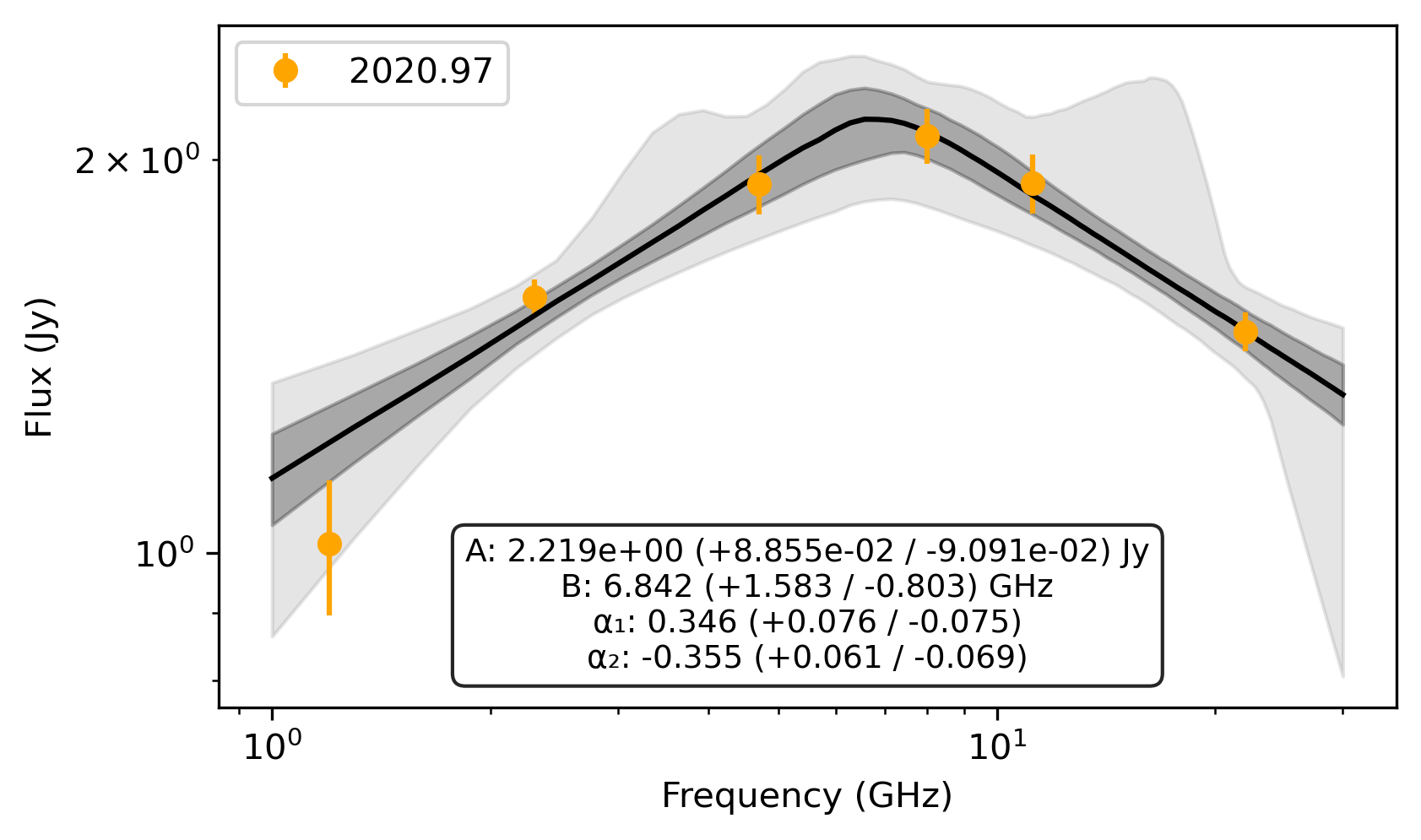}
    \end{subfigure}

    \caption{\emph{Left Panel:} Evolution of the spectral break and power-law slopes from the RATAN radio lightcurve \cite{2024MNRAS.527.8784A}. \emph{Right Panels} For clarity, we plot with fainter points all epochs where the uncertainties of the slopes were larger than 0.2.  Two characteristic fits of radio spectra from individual epochs are presented.}
    \label{fig:ratan_break}
\end{figure}

RATAN data provide lightcurves in 6 frequencies between 1.2-22 GHz. To investigate the radio spectral shape during the outburst, we fitted individual epochs with a smooth broken power-law model to obtain parameters, such as the break frequency and slopes before and after the break: 

\begin{equation}
F(x) = A_0 \left( \frac{x}{x_\mathrm{b}} \right)^{\alpha_1} 
\left[ 1 - \frac{1}{2} \left( 1 + \tanh\left( \frac{x - x_\mathrm{b}}{\delta_{sm}\, x_\mathrm{b}} \right) \right) \right] 
+ A_0 \left( \frac{x}{x_\mathrm{b}} \right)^{\alpha_2} 
\frac{1}{2} \left( 1 + \tanh\left( \frac{x - x_\mathrm{b}}{\delta_{sm}\, x_\mathrm{b}} \right) \right),{\rm and\ } \delta_{sm}=0.1,
\label{eq:smooth}
\end{equation}

where, $F(x)$ is the flux density at the observed frequency $x$, $A_0$ is the amplitude normalization at the break $x_\mathrm{b}$, $\alpha_1$ and $\alpha_2$ are the power-law indices below and above the break, respectively, and $\delta_{sm}$ controls the smoothness of the transition, with smaller values corresponding to a sharper break. For the fit we adopted $\delta_{sm}=0.1$ to limit free parameters.
We fitted each epoch with at least 4 RATAN measurements with the above mentioned model using a Bayesian approach. We derive posterior probability distributions and the Bayesian
evidence with the nested sampling Monte Carlo algorithm
MLFriends \citep{2016S&C....26..383B,2019PASP..131j8005B} using the
UltraNest\footnote{\url{https://johannesbuchner.github.io/UltraNest/}} package \citep{2021JOSS....6.3001B}. 
For the priors, we used uniform distributions for all parameters,
with $\alpha_1, \alpha_2 \in [-3, 3]$, $x_{\mathrm{b}} \in [1, 22]\,\mathrm{GHz}$ and $A_0 \in [0,4]$ Jy.
The results are shown in Fig. \ref{fig:ratan_break}, where in the left panel we plot the derived best fitted parameters during the outburst, and in the right panel we show the fit during two individual epochs. It is clear that the break remains somewhat constant, while there is a scatter, there is an indication that it moves towards lower frequencies at the decay of the outburst. However the main takeaway is that the slope before the spectral break remain in the range $0\lesssim \alpha_1 \lesssim 1$  (see top panel in the left column of  Fig.~\ref{fig:ratan_break}) which is substantially softer than the optically thick SSA expectation of $F_{\rm \nu}\propto \nu^{5/2}$ \citep{1986rpa..book.....R}.

\bibliographystyle{aasjournalv7}
\bibliography{ref}
\end{document}